\documentclass[a4paper,oneside,12pt]{article}
\usepackage[T1]{fontenc}
\usepackage[latin1]{inputenc}
\usepackage[english]{babel}

\usepackage{lmodern} 
\usepackage{authblk}
\usepackage{url}

\usepackage{graphicx} 

\usepackage{amsmath}
\usepackage{amsthm}
\usepackage{amsfonts}

\usepackage{ulem} 
\usepackage[usenames]{color}
\begin{document}


\title{Symbolic Dynamics of Music from Europe and Japan}




\author[1]{Vasileios Basios}

\author[2]{Robin De Gernier}

\author[3]{Thomas Oikonomou}

\affil[1]{Université Libre de Bruxelles, Interdisciplinary Center for Nonlinear Phenomena 
and Complex Systems \& Service de Physique des Systémes Complexes et Mécanique 
Statistique, BruxellesBelgium}

\affil[2]{OPERA - Photonique, Faculté des Sciences appliquées-école polytechnique, Physique et 
Mathématique}

\affil[3]{Department of Physics, Nazarbayev University, 53 Kabanbay Batyr Avenue, Astana 010000, 
Kazakhstan}

\date{\today}                     
\setcounter{Maxaffil}{0}
\renewcommand\Affilfont{\itshape\small}

\maketitle

\begin{abstract}

After a brief introduction to the theory underlying block-entropy, and its relation to the dynamics  
of complex systems as well as certain information theory aspects, we study musical texts 
coming from two distinct musical traditions (Japanese and Western European) encoded via 
symbolic dynamics. We quantify their information content or also known as the degree of 
``non-randomness'' which essentially defines the complexity of the text. We analyze  the departure 
of ``total randomness'' to the constrains underlying the dynamics of the symbol generating process. 
Following  Shannon on his attribution to these constraints as the emergence of complexity, we 
observe that it can be accurately assessed by the texts' block-entropy versus block-length scaling 
laws.
\end{abstract}


\pagestyle{plain} 




\newpage

\section{Introduction}
One of the first accounts of a learned westerner \cite{Piggott} when encountering 
the rich Japanese musical tradition highlighted the great differences in genre, tonality and 
methodological ethos between western and eastern music, expressing a great admiration toward it. As 
it is well known, the Chinese and Japanese tuning also differs from the established western 
equal-tempered scale. In this work we put forth a first, to our knowledge, comparative analysis of 
two classic music texts from the European and the Japanese tradition, using the tools and concepts 
stemming from complex system's science \cite{Nicolis and Nicolis 2007, Hao, Kitchens}.

Many natural or man-made processes can be recorded in the form of a sequence of symbols, 
which we refer to as ``text''. This text is information-rich and as such can be studied in a way 
similar to that of complex systems in the framework of Symbolic Dynamics.
Already from the early studies in the area of information dynamics of complex nonlinear systems the 
analysis of musical texts using the conceptual and computational tools  of Statistical Mechanics 
were pursued \cite{Nicolis1991, Boon1995, NicBasBook}. The same tools have been used in a variety 
of information rich texts, either natural or man-made; for characteristic publication on the 
subject one can see for example \cite{Kalimeri2012, Martinakova2008} for literature texts and 
music, \cite{Berthe1994} for sequences generated by automata and \cite{Oikonomou2007} for DNA 
coding and non-coding sequences.

In this study, after a proper definition of information and useful concepts such as Shannon 
block-entropy, we compare music texts coming from the two distant musical traditions, 
the Japanese and the Western European ones:
(a) From the Western European we have chosen a representative and well studied piece of 
classical western music, Beethoven's \textit{Sonata Op.31 No.2} \cite{Beethoven} and 
(b) from the Japanese musical tradition two traditional equally classic and well studied 
music pieces, Yatsuhashi Kengyoo's \textit{Rokudan no Shirabe} \cite{Yatsuhashi Kengyoo} and 
Yoshizawa Kengyoo II's \textit{Chidori no kyoku} \cite{Yoshizawa Kengyoo II}. The latter author's 
life and work (circa 1808-1872) overlaps with the period of Beethoven's life and work (1770-1827).

A naturally arising issue is whether or not these different musical compositional genres and 
their syntax and semantics would show any quantifiable differences under the scrutiny of the tools 
and methods already in use in complex systems. Since the musical text, or for that matter any text, 
can be viewed as a coded, one-dimensional, symmetry-broken, unfolding trace of the complex 
dynamical process that produced it.

We ask ourselves the question ``Do the audible differences of the music pieces reflect in the 
block-entropy analysis?'', or in a more simple way ``Does one observe differences between music 
pieces in the block-entropy analysis?''. 
As the answer to this question turns to be affirmative we conclude this work by looking for the specific power laws that govern the scaling  of the block-entropy of these music texts. For references and comparison we also study some well known dynamical maps and present a few selected results relevant to our study.

The structure of the paper is as follows: Section \ref{MaM} provides the key conceptual 
and computational tools from complex systems and symbolic dynamics utilized herein. 
These include the methods of encoding and informational processing of musical pieces,
introducing the concept of discrete Shannon, or Block, Entropy, and the measure of `uncertainty per 
word' as well and the finite length effects present in any realistic implementation of these 
mathematical concepts. Subsequently, in Section \ref{Sec3}, we investigate dynamical maps on the interval with 
well defined entropic characteristics in order to review the entropic concepts introduced before. 
Then the analysis concerning the musical texts is presented and their characteristic times and 
scaling laws are identified and discussed. In the final section we conclude and discuss the possible 
outlook and relevance of this work.
\newpage

\section{Materials and methods}\label{MaM}

The present work being mostly made of computations, the materials used are of course not of the 
physical, tangible type. A major part of these materials is software. In this section we 
could describe the detailed functioning of the programs used but that would only bring tiredness to 
the reader. We will only mention those programs while exposing the method of 
work applied, and highlight their main functions. Links and credits to the programs can be found in  the References at the end of this work. The last part of the current section is dedicated to a short introduction to the block-entropy analysis and related theory, followed by illustration example.

\subsection{\textit{Encoding and processing musical pieces}}
A most important, and tangible - at the very least audible -, material of this study are the scores 
 of the chosen musical pieces. The scores used were in the form of {\tt .midi} files. This format 
cannot be  directly used for our analysis and needed first to be converted in human-readable format. 
This was  done by the software {\tt MIDIFile2Text} \cite{Byrd} which converted {\tt.midi} files 
into .text files  containing various information such as the time the note was played, its duration 
and its tone  which were the variables used for our study. The file conversion process  {\tt 
MIDIFile2Text} also  separated the tracks/instruments of the {\tt .midi} files which proved  quite 
helpful given  that we are mainly looking for monotonic samples.

The method used to analyse the music samples was based on the three letters encoding 
approach of Ebeling \& Nicolis on Beethoven's \textit{Sonata Op. 31 No.2} 
\cite{ebenicolisa}, slightly adapted in an attempt to better conserve the 
dynamics of the music. The three symbols or `letters' used here for the encoding of the 
musical text are:
\begin{itemize}
	\item[-] $u$ for up, when the tone of the note rises in comparison to the previous one
	\item[-] $d$ for down, when the tone of the note drops in comparison to the previous 
one
	\item[-] $s$ for sustains, when the tone of the note remains the same in comparison 
to the previous one.
\end{itemize}

In the study of the dynamic of the music the duration of notes is possibly an important variable. 
For example, this is one of the main audible differences between occidental and Japanese 
traditional music \cite{world_music}. The duration of notes in Japanese pieces tends to be much 
longer than in occidental pieces \cite{music_interaction}. Will this difference show itself in the 
block-entropy analysis? This is one of the questions we asked in the introduction and will try to 
answer in the following.

The challenging step for the block-entropy procedure was how to describe the duration of  a single 
note by means of the three lettered symbolic representation.
The natural solution is to repeat the letter ``$s$'' a number of times 
proportional to the duration of the note. This requires to chose a time unit, a time scale, a choice 
which will probably impact strongly on the dynamics of the music. The unit should be common to all 
of the pieces: we choose the smallest measure of time present among our pieces. Another way of 
understanding this method is to see the music as a succession of notes with a constant duration 
equal to the chosen time unit. In this view, a sustained note is in fact a succession of shorter 
notes of the same tone, the number of which is given by the division of the global duration by the 
time unit; and thus a natural time-scale is defined.

Once the encoding was constructed the resulting text is processed by a program in {\tt C++} 
provided by one of the authors and freely available in a public code repository \cite{entropa}.
In the course of this study we calculated all the probabilities (i.e. relative frequencies), the 
Shannon block-entropy and the Shannon entropy of words ranging from $2$ letters (the minimum 
block-length) to $32$  letters (the maximum block-length due to our computing time limitation) 
letters.

\subsection{\textit{Methods of comparison and analysis}}

The second part of the current section concerns the way we analyzed and compared our musical pieces 
to well-known discrete dynamical systems on the interval, i.e. dynamical maps such as the Logistic 
Map.

Each of these maps, via a suitable coarse-graining of its state-space \cite{Nicolis and 
Nicolis 2007, Basios et al 2008, Basios 2008b} produces a ``text'' in a binary alphabet which is 
then processed by the program computing the entropies \cite{entropa}, in the exact same way as for 
the musical ``texts''.  The comparison 
is then made on basis of the graphs of the block-entropy against the length of the words.
Here is a list of the maps with a short description for each of them:

\begin{itemize}
\item[-]	Manneville's intermittent systems :
\begin{equation}\label{dyneq1}
x_{n+1} = x_n + x^{z}_{n} \quad \text{(mod 1)}\,.
\end{equation}
Those systems are known to be intermittent for $z < 2$ and sporadic when $z \geq 2$ \cite{Gaspard 
and Wang 1987}. 
We studied them with values $z = 0.5$ and $z = 2$.
\end{itemize}

\begin{itemize}
\item[-] A map used in \cite{Basios et al 2008}: 
\begin{equation}\label{dyneq2}
f(x)=\begin{cases}
    f_L(x)=\frac{a-\sqrt{a^2-(4a-2)x}}{2a-1} & \text{if } 0\leq x<\frac{1}{2}\\
    f_R(x)=f_L(1-x)       & \text{if } \frac{1}{2}\leq x<1
\end{cases}\,.
\end{equation}
This map is known to be intermittent in the case $a = 1$, which is the case studied here. 

\end{itemize}

The study of Eqs. (\ref{dyneq1}) and (\ref{dyneq2}) is presented in Section \ref{Sec3} for $z=\{0.5,2\}$ and $a=1.0$, respectively, and their results of the Block Entropy analysis are recorded in Fig. \ref{Fig04}.

\subsection{\textit{About block-entropy}}
It is no original approach to give here a brief introduction to the theory underlying block-entropy 
and its relation to dynamical systems. As a matter of fact this is a pattern tirelessly repeated 
over the numerous works, articles, books\ldots which can be found on the subjects of complex 
systems, information theory or block-entropy analysis itself.  Therefore the reader could find all 
he needs (and wishes) to know about block-entropy by consulting the references given at the end of 
this work (in particular, see \cite{Hao, Basios et al 2008, Basios 2008b,Nicolis 2005} for a 
comprehensive review).
However, on behalf of the self-consistence of the present work, it cannot be avoided mentioning it.

Many natural (physical, biological) or man-made processes (mathematical sequences, social processes, 
written texts and/or musical scores) can be recorded, or ``written down'', in a symbol sequence. 
This symbol sequence -or ``text''- is essentially one-dimensional, information-rich and 
uni-directional, in space as well as in time. The information content, or else the degree of 
``non-randomness'', is essentially what defines the complexity of the text. After the seminal work 
of Shannon in the 1950's a plethora of methods have been proposed in order to quantify this 
complexity. Shannon himself, attributed the emergence of complexity, i.e. the departure of ``total 
randomness'' to the constrains underlying the dynamics of the symbol generating process 
\cite{Shannon}.

A measure of the dependence of the number of sequences of a given length (to which we will refer 
hereafter as ``words'') with respect to the length is a Shannon-like entropy measure, the so called 
``block-entropy''. Let us define this notion explicitly \cite{Nicolis and Nicolis 2007,Hao, Kitchens}. We consider a text composed on a certain 
alphabet $\{A_1,\ldots,A_m\}$ and which can be divided in words of length $n$.
The block-entropy is given by
\begin{equation}\label{Hn}
H_n = -\sum p^{(n)}(A_1 \cdots A_n)\log p^{(n)}(A_1 \cdots A_n)\,,
\end{equation}
where the $p^{(n)}(A_1 \cdots A_n)$ are the probabilities (relative frequencies) associated 
with the words $(A_1 \cdots A_n)$. The summation is made over all the words ``$A_1 \cdots A_n$'' 
and it is understood that $0\log 0=0$. Note that for uniformly distributed probability, 
block-entropy is exactly the logarithm of the number of words of a given length $n$.

The following notions are also essential for the course of this study.
The uncertainty of the next letter appearing after a given block of a given length $n\geq 1$:
\begin{equation}\label{defhn}
h_n=H_{n+1}-H_n
\end{equation}
with $H_0=0$.
Then, the entropy of the source which is a discrete analog of the Kolmogorov-Sinai entropy and 
is defined as:
\begin{equation}\label{defh}
h=\lim_{n\to \infty}h_n\,.
\end{equation}
The entropy of the source can be shown, for a given alphabet, to take the largest value for a 
Bernoulli processor \cite{Hao,Gaspard 1998}. Evidently, for \textit{m}th order Markov 
sources \cite{Kitchens} the limit in Eq. (\ref{defh}) is reached for $n=m$ i.e. $ h_n=h$ if $ n\geq 
m $.

For sequences generated by Bernoulli or Markov processes the Shannon-McMillan theorem  \cite{Applebaum 1996} asserts that the number of words of length $n$ in Eq.(\ref{Hn}) scales exponentially with $h$. Recalling from before that for uniformly distributed probability the block-entropy is exactly the logarithm of the number of words of length $n$, this implies that:
\begin{itemize}
\item[(i)] block-entropy scales linearly with the word length (Eq. (\ref{Nlin_n})),
\item[(ii)] long words are extremely improbable, being exponentially penalized (with a maximal penalization for the largest value of $h$, that is for a Bernoulli processor).
\end{itemize}

This being said, many real world systems are not Bernoulli or Markovian processors and we do not 
observe exponential rates of growth of the number of words of length $n$. Therefore there must exist 
a procedure of selection of realizable sequences \cite{Basios et al 2008, Nicolis 2005}. In this 
context we can assume \cite{ebenicolisa} that for a large class of systems the following scaling 
behavior is valid for the block-entropy:
\begin{equation}\label{ScLaw}
H_n = nh+gn^µ(log n)^\nu + e
\end{equation}
with $0< µ <1$ and $\nu<0 \notag$, where $h$ is the entropy of the source defined above. Ebeling and 
Nicolis showed that for classical literature texts or music the block-entropy parameter values are 
$h=0$, $\nu =0$ and $µ<1$. Dynamical systems showing weak chaos in the form of intermittency, and 
sporadic systems have also been shown to give rise to a sub-linear scaling of their block-entropy 
\cite{Nicolis and Nicolis 2007, Gaspard 1998}).

\subsection{\textit{Illustration on an example}}
The illustration example chosen comes from \cite{ebenicolisa}. The purpose here is not only 
to illustrate our definitions above, it is also to demonstrate our ability to generate and process 
``texts'' from a dynamical source, therefore demonstrating the good functioning and use of the 
programs mentioned above. We will not interpret the results on this example (the 
interpretation can be found in the work of Ebeling \& Nicolis \cite{ebenicolisa}).

We study the tent map with $1<r<2$:
\begin{equation}\label{TentMap}
f(x)=\begin{cases}
    rx & \text{if } 0< x<\frac{1}{2}\\
    r(1-x) & \text{if } \frac{1}{2}< x <1
\end{cases}
\end{equation}
This map generates a text (we chose a length of $10^4$ letters\footnote{The length of the texts in 
this study will always be $10^4$ letters.} on the alphabet $O$ and $L$, given the partition: 
$O \leftarrow x\in(0,1/2)$, $L \leftarrow x\in(1/2,1)$. In our study we consider the four cases 
$r=\frac{1}{2}(1+\sqrt{5})\approx 1.618$ and $r=2^{{1/2}^k}$ with $k=1,2,3$.

\begin{figure}[!ht]
\centering
\includegraphics[keepaspectratio,width=8cm]{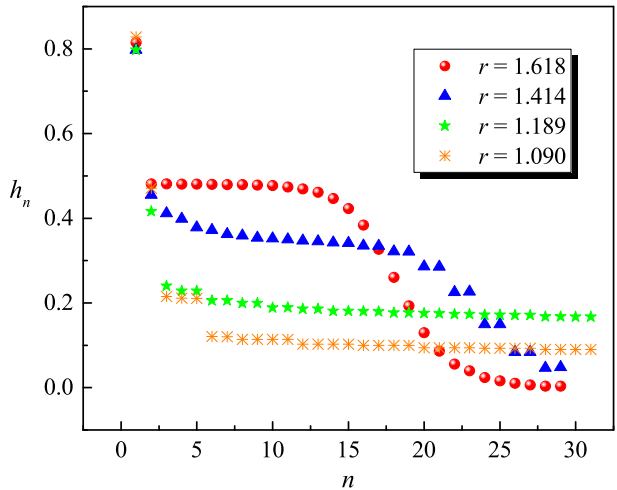}
\caption{Uncertainty $h_n$ for the tent map vs. the length $n$ of the words for several values of the parameter $r$.}
\label{tentmapf}
\end{figure}

As we can see on Fig. \ref{tentmapf}, we do obtain the same results of Nicolis and Ebeling  \cite{ebenicolisa}. We have deliberately pushed the study of the system to values of \textit{n} on a  wider range than Nicolis and Ebeling ($0 \to 30$ in place of $0 \to 10$) so as to show the effects  of finite length of the text on the block-entropy. One can see on Figure \ref{tentmapf} that the  uncertainty per letter $h_n$ goes to $0$ for $n > 12$. This will be the subject of the following  section. 

\newpage
\subsection{\textit{Finite length effects on block-entropy}}
As it has been exposed precedently, when no analytic derivation of the probabilities can be made, as it is the case for texts generated by language-like processes, those probabilities can be estimated by the frequencies of words. It appears quite clearly that this estimation will gain in precision with the total length (in terms of number of letters) of the text. In particular, for words with large length $n$ which may not be given the chance to appear if the text is too short. The required length to obtain a good approximation of the probabilities depends thus on the length of the words, but also on the selectivity in words of the dynamics generating the text. Were all words of length $n$ equiprobable, the risk of them not appearing frequently enough would be greater. 
To realize this, consider first the extreme case where only one word is allowed. It is clear that this will give the right probability (equal to $1$). 
Now if one word is allowed with great probability (say $0.80$) while several others are allowed with smaller probabilities, the chance is low 
that the most probable word does not appear frequently enough as to give a correct estimate. The probabilities of the other being small, their contribution to the block-entropy will be small as well and thus the estimate of the block-entropy will be correct.
We can express the condition on the total length of the text in a more mathematical way. We take $N_n$ as the total number of allowed words of length $n$, $L$ the total length of the text. The condition is:
\begin{equation}\label{Nlin_n}
N_n\ll L\,.
\end{equation}

In the case of equiprobable words of length $n$ (that is a generator of Bernoulli type) on an  alphabet of $\alpha$ letters:
\begin{equation}\label{NtoA}
N_n=\alpha^n
\end{equation}
which leads us to the condition $n\ll \log_\alpha L$.
If we consider more selective dynamics, \cite{ebenicolisa, Basios et al 2008, Nicolis 2005}, we may 
have:
\begin{equation}\label{NtoSubEx}
N_n=\alpha^{\sqrt{n}}
\end{equation}
which gives the more permissive condition $\sqrt{n}<<L$.

The example of the tent map given in the previous section, through its comparison with the study of 
Nicolis and Ebeling, demonstrates perfectly the effects of the finite length $L$ on the 
block-entropy at large values of $n$. This system, for $r=1.618$, being the generator of a first 
order Markov chain, the uncertainty per letter, Eq. (\ref{defhn}), of the source is 
analytically known to be $h=0.694$, which is verified numerically for $n\leq 10$ as shown on figure 
\ref{tentmapf}. However, it goes to zero for $n>12$ when it should remain constant. This indicates 
that, due to the finite length $ L $, no more information is added for large $n$.
One can also see on figure \ref{tentmapf} that those effects of finite length appears for greater $n$ for the other values of the parameter $r$. This is an illustration of the fact that the equiprobable distribution is the most penalized in terms of precision of the estimation.

\begin{figure}[!ht]
\centering
\includegraphics[keepaspectratio,width=8cm]{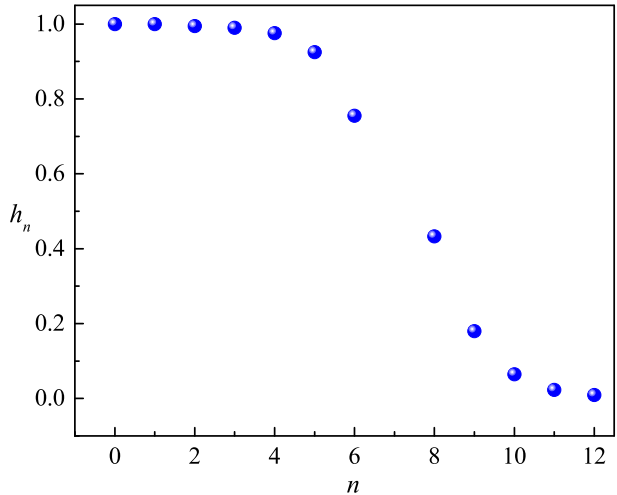}
\caption{Uncertainty $h_n$ for a 3-letters Bernoulli Generator vs. the word length $n$.}
\label{Fig02}
\end{figure}

Another example of the latter statement is presented in Fig. \ref{Fig02}, this time with a 
3-letters Bernoulli Generator and a text of length $L=10^4$. We see that the estimate is correct 
(cf. Eq. (\ref{Nlin_n})) only for $n<\log_{3}(10^4)$, i.e. approximately $n<8$.

\newpage

\section{Results and discussion}\label{Sec3}

In this section we compare musical pieces and/or dynamical maps, including of course a  discussion 
on the scaling laws. 
We begin by exposing the results for the dynamical maps described by Eqs. (\ref{dyneq1}) and 
(\ref{dyneq2}).

\subsection{\textit{Dynamical maps}}

Both maps in Eqs. (\ref{dyneq1}) and (\ref{dyneq2}) are studied on the interval $[0,1]$ on a 
2-letter (binary) alphabet defined by the usual partition: $O \leftarrow x\in 
(0,1/2)$, $L \leftarrow x\in(1/2,0)$. The texts generated were of $10^4$ letters long.

\begin{figure}[!ht]
\centering
\includegraphics[keepaspectratio,width=8cm]{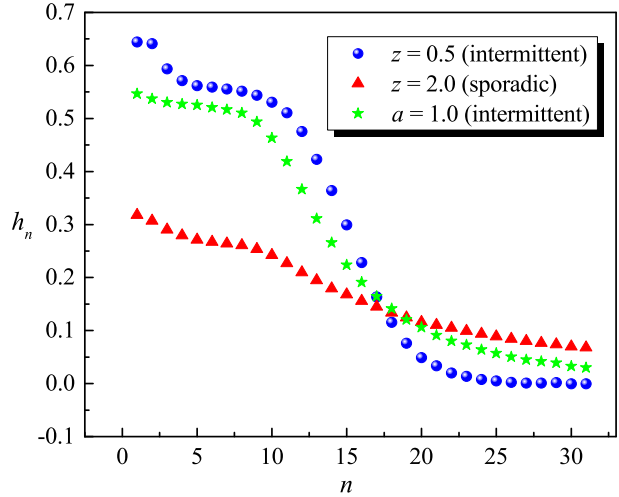}
\caption{Uncertainty $h_n$ vs. the word length $n$ for the two maps defined  by Eqs. (\ref{dyneq1}) (parameter $z$) and (\ref{dyneq2}) (parameter $a$) for $L=10^4$.} 
\label{Fig03}
\end{figure}

\begin{figure}[!ht]
\centering
\includegraphics[keepaspectratio,width=8cm]{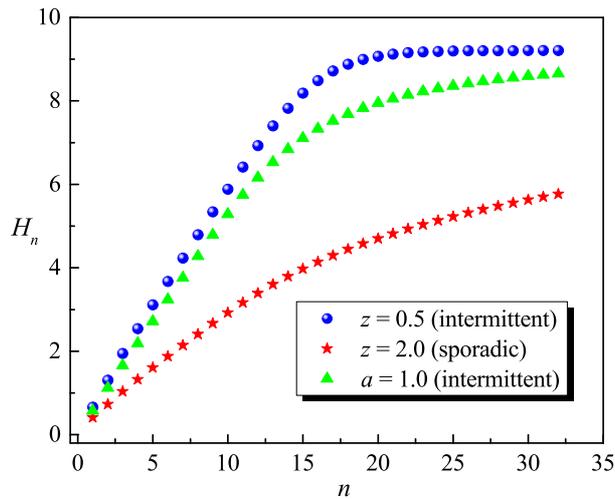}
\caption{Block-entropy $H_n$ (in bits) vs. the length of the words $n$ for the two maps defined by Eqs. (\ref{dyneq1}) (parameter $z$) and (\ref{dyneq2}) (parameter $a$) for $L=10^4$.}
\label{Fig04}
\end{figure}

One can see on Figs. \ref{Fig03} and \ref{Fig04} that the two intermittent systems behave very similarly, while 
the sporadic system adopts a quite different aspect. The curves of the latter, both for $H_n$ and 
$h_n$, are indeed way smoother than those of the intermittent systems and take on overall smaller 
values. This can be explained by the fact that a sporadic system spends more time around a specific 
point where its derivative is null, thus favouring the dominance of one letter of the alphabet on 
the other and therefore reducing the gain of information (that is of entropy, in the sense of 
Shannon) by addition of a letter. Another way to see this is that for a sporadic system, the 
Lyapunov exponent vanishes \cite{Gaspard and Wang 1987}, giving a greater stability to the system.

\newpage
\subsection{\textit{Musical analysis}}

\begin{figure}[!ht]
\centering
\includegraphics[keepaspectratio,width=8cm]{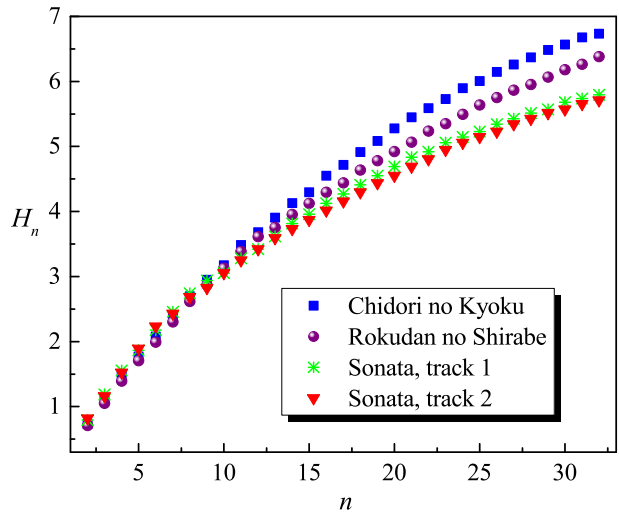}
\caption{Block-entropy $H_n$ (in bits) vs. word length $n$. Blue squares ``Chidori no kyoku'', purple dots ``Rokudan no Shirabe'', green crosses ``Sonata for pianoforte Op. 31 No.2, track 1'', red triangles ``Sonata for pianoforte Op. 31 No.2, track 2''.}
\label{Fig05}
\end{figure}

\begin{figure}[!ht]
\centering
\includegraphics[keepaspectratio,width=8cm]{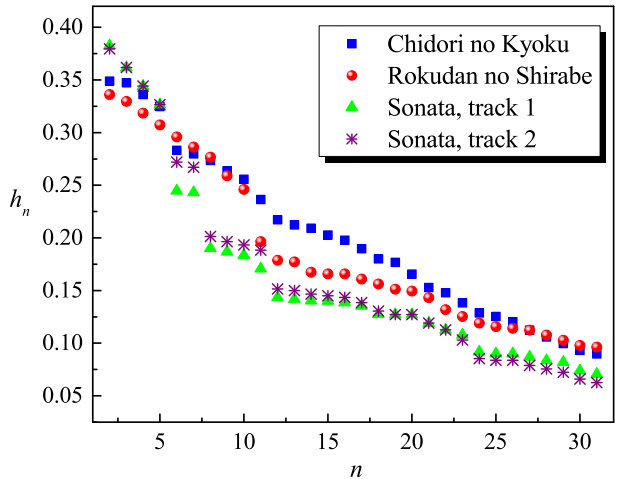}
\caption{Uncertainty $h_n$ (in bits) vs. the word length $n$. Blue squares ``Chidori no kyoku'', red spheres ``Rokudan no Shirabe'', green triangles ``Sonata for pianoforte Op. 31 No.2, track 1'', purple crosses ``Sonata for pianoforte Op. 31 No.2, track 2''.}
\label{Fig06}
\end{figure}

The method used to analyze the musical pieces has been largely exposed in the Section \ref{MaM}. With our method of transcription, we obtained four texts of the following  lengths: ``Rokudan no Shirabe'' $L=6644$, ``Chidori no kyoku'' $L=7909$, ``Beethoven track 1''  $L=11616$, ``Beethoven track 2'' $L=11968$. We can concentrate ourself on the interpretation of the  Figs. \ref{Fig05} and \ref{Fig06}. They resume the results of the four pieces for the  block-entropy and the uncertainty per letter, respectively.

The first thing to be noticed on these two graphs is the encouraging similarity between the curves 
of the two tracks from the Sonata of Beethoven, as well as the dissimilarity of the letters towards 
the Japanese pieces. This is easier seen on the Fig. \ref{Fig06}. Beethoven's tracks follow exactly 
the same ``staircase'' pattern, with a small shift towards the y-axis. They even coincide with great 
precision for $18\leq n \leq 22$. This result in itself is correct and encouraging because it shows 
that our block-entropy analysis is consistent with the fact that both tracks came from the same 
piece of music and are therefore generated by the same dynamics, despite their audible small 
differences. That is something we expected as a reliable verification and we are glad to 
have been observed.
The two Japanese pieces are quite shifted from Beethoven's tracks, close to each other for $n<11$ 
and far apart for $11\leq n<21$. Again it is reassuring to see the differences between text from 
distinct origins and it can give us faith in our method. However, this leads to small values of 
$H_n$ and $h_n$, which is observed in our figures. Those small values are a first point of analogy 
with the sporadic system given by the Manneville's systems for $z=2$. Of course, this is subject to 
discussion as the analogy seems to be the direct consequence of our method of encoding, but it is 
well known that block-entropy analysis depends on the chosen alphabet/encoding.

Let us now comment the ``staircase'' structure appearing on figure $6$, in particular for the uncertainty per letter $h_n$ of both Beethoven's tracks. This is related to the time unit we chose to generate our text. The values of $n$ at which a new step occurs correspond indeed to multiples of this time unit. The fact that our analysis allows the extraction of information such as characteristic times of the musical pieces is quite satisfactory in itself!

\begin{figure}[!ht]
\centering
\includegraphics[keepaspectratio,width=8cm]{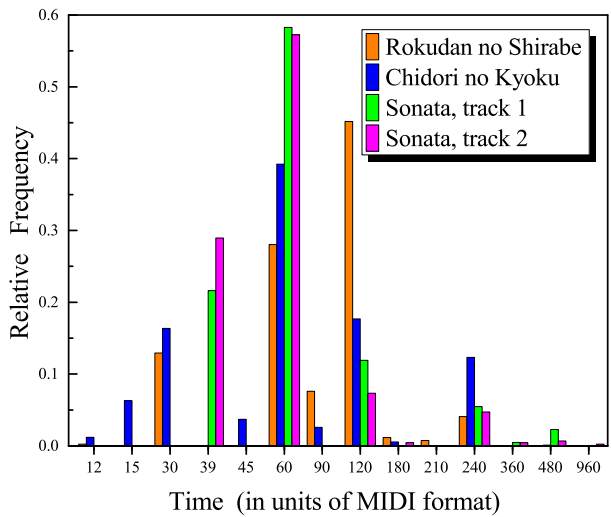}
\caption{Characteristic times in units of the .MIDI files.}
\label{Fig07}
\end{figure}

We illustrate this in Fig. \ref{Fig07} which shows the characteristic times (and their frequency) of the three musical pieces in units given by the .MIDI files. The time unit we have chosen is the smallest appearing time, i.e. 12. Among the characteristic times in Fig. \ref{Fig07}, some are (close to) multiple of this time unit, the ratios being given by: $n=5$ for $T=60$, $n=7,8$ for $T=90$, $n=10$ for $T=120$ and $n=20$ for $T=240$. When comparing Figs. \ref{Fig06} and \ref{Fig07} one sees that the steps for a given piece of music coincide indeed with the characteristic times of that piece shown in Fig. \ref{Fig07}.

Fig. \ref{Fig06} thus reflects the fact that Beethoven's Sonata has shorter characteristic times and 
less varied while Japanese music is characterized by longer and more diverse times. The answer to 
the question ``Will the difference in duration of notes show itself in the block-entropy analysis 
?'' asked in Section \ref{MaM} is then \textit{yes, it does}!

\subsection{Scaling laws}

In this section, we discuss the scaling laws of the pieces of music; cf. Eqs. 
(\ref{Nlin_n}, \ref{NtoA}, \ref{NtoSubEx}). Our objective is to show that the scaling behavior of 
Eq. (\ref{ScLaw}) is valid with $h=0$ and some value of $\mu$ and $\nu$.

\begin{figure}[!ht]
	\centering
	\includegraphics[keepaspectratio,width=8cm]{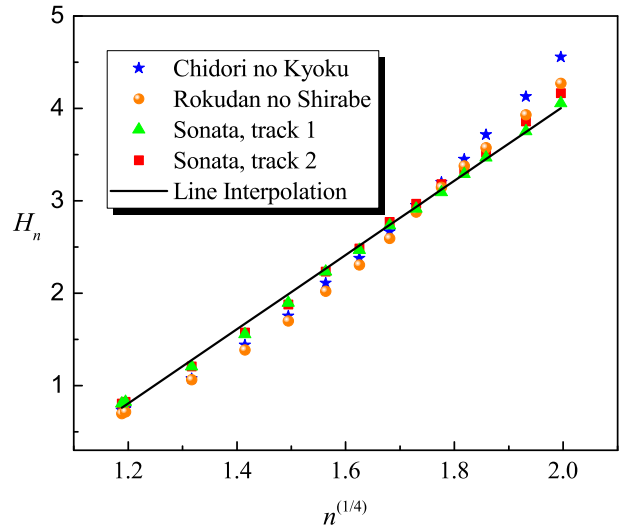}
	\caption{Power law in $n^{(1/4)}$ of the block-entropy $H_n$ (in units of $\log_3$) for Beethoven's Sonata.}
	\label{Fig08}
\end{figure}

For what the pieces of music are concerned, Fig. \ref{Fig08} (where the entropy is represented 
against $n^{(1/4)}$ and where a linear fit has been made) shows that the entropy of Beethoven's 
Sonata follows a power law with $\mu=1/4$, which is the same result as in Ref. 
\cite{ebenicolisa}. 
The Japanese pieces however do not follow the same law, though it may be difficult to 
easily see it on Fig. \ref{Fig08}. 

\begin{figure}[!ht]
	\centering
	\includegraphics[keepaspectratio,width=8cm]{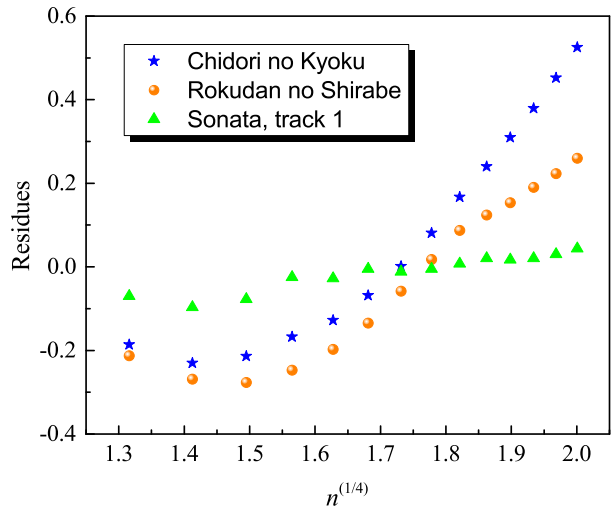}
	\caption{Distance of the curves to the linear fit in Fig. \ref{Fig08}.}
	\label{Fig09}
\end{figure}

The deviations from the linear fit were represented on Fig. \ref{Fig09} where it is easier seen that 
only Beethoven's Sonata follows a power law with $\mu=1/4$, $\nu =0$. The track 2 of Beethoven's  
Sonata is not represented in order to avoid redundancy. Again, the observation of this difference  
between the Japanese pieces and the Sonata is in agreement with our expectations that these pieces  
of music correspond to distinct dynamics.

\begin{figure}[!ht]
\centering
\includegraphics[keepaspectratio,width=8cm]{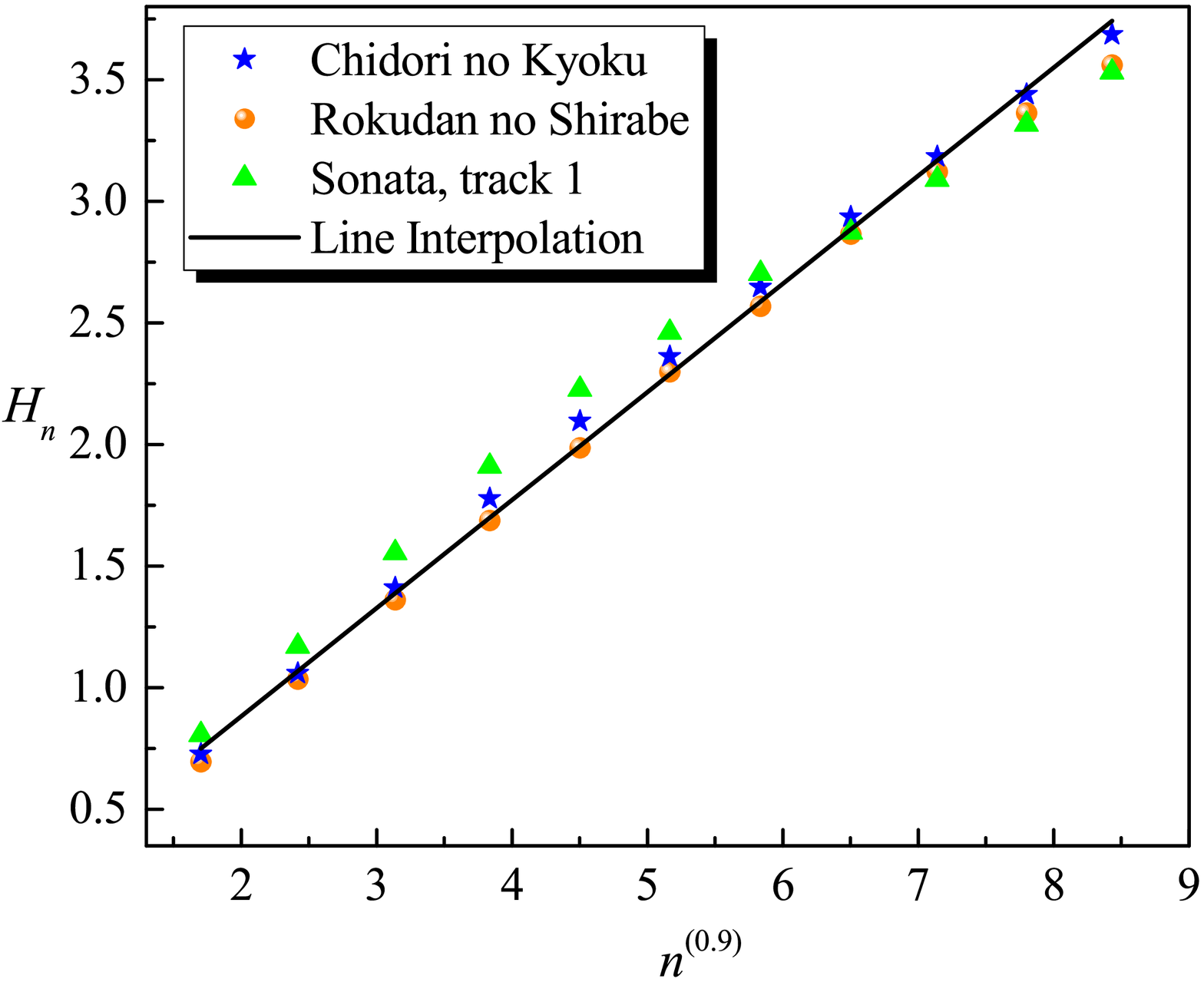}
\caption{Power law in $n^{(0.9)}$ of the block-entropy $H_n$ (in units of $\log_3$) for 
\textit{``Rokudan no Shirabe''}.}
\label{Fig10}
\end{figure}

\begin{figure}[!ht]
\centering
\includegraphics[keepaspectratio,width=8cm]{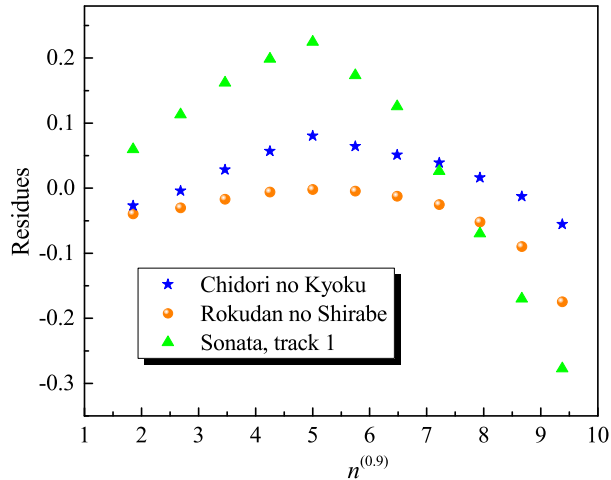}
\caption{Distance of the curves to the linear fit in Fig. \ref{Fig10}.}
\label{Fig11}
\end{figure}

Similarly, we can repeat the process for \textit{Rokudan no Shirabe}. We find that its block-entropy scales in a power law with $\mu=0.9$. This is illustrated on Fig. \ref{Fig10}. The residues are represented on Fig. \ref{Fig11} for easier visualisation. We see that block-entropy of the other pieces do not fit to the same power law.

\section{Conclusion}

In the course of this study, after having properly defined the concepts and tools in use 
(information, block-entropy, entropy of the source, \ldots), we have been able to apply them on 
various systems: dynamical, mathematical maps, western music text and Japanese music texts. 

The study of the dynamical maps defined by Eqs. (\ref{dyneq1}, \ref{dyneq2}) and 
(\ref{TentMap}) allowed us to observe the differences between fully chaotic, intermittent and 
sporadic systems. We noted that sporadic systems present smoother block-entropy $H_n$ and 
uncertainty per letter $h_n$ curves with significantly lower values than intermittent systems.

This result, in regard of the curves of the block-entropy and uncertainty per letter for the music 
texts gave us insights on the nature of the dynamics generating those music pieces. Furthermore, we 
showed that both Japanese and occidental chosen pieces have their block-entropy scaling in a 
sub-linear way (Eq.(\ref{NtoA})) with the words length. In particular Yatsuhashi Kengyoo's 
\textit{Rokudan no Shirabe} 
and Beethoven's \textit{Piano Sonata Op.31 No.2} were shown to have a power-law block-entropy 
scaling with distinct exponents which, as we saw, were $\mu=1/4$ for \textit{Piano Sonata Op.31 
No.2} (a new result yet consistent with the general theory developed in the first 
study of this piece by Ebeling and Nicolis) and $\mu=0.9$ for \textit{Rokudan no Shirabe}. We 
therefore obtained a first favorable answer to the introductory 
question ``Does one observe differences between music pieces in the block-entropy analysis?''.

Before studying the block-entropy scaling we conducted a first analysis of curves of the 
block-entropy and uncertainty per letter of the music pieces, carefully indexing the differences and 
similarities between them. We observed a great similarity between the two tracks of Beethoven's 
\textit{Piano Sonata Op.31 No.2} and a relative similarity between the Japanese pieces for words of 
length inferior to $n=11$, thus validating our method of analysis. Moreover, we noted 
``staircase-like'' structure in the uncertainty per letter curves and identified them with the 
presence (or absence for some music pieces) of characteristic times, giving rise to the differences 
between the curves. We therefore answered the introductory question ``Do the audible differences of 
the music pieces reflect in the block-entropy analysis?''. The answer is yes, they do.

In later work, we might address the question of the dependence in the time-scale (or in the choice 
of the time unit) of the block-entropy and entropy per letter of the music pieces. We expect this 
dependence to be important due the link between characteristic times of the music and entropy per 
letter.
It would be also very interesting to look for the presence  of palindromes in the generated music texts, which can directly be accomplished using 
\cite{entropa}. This question has recently attracted some interest \cite{Bach-Recurrence} as it is 
 related to both to recurrence quantification analysis and also to entropic analysis. Any future 
connection between these two approaches will prove to be an excellent step toward a more general 
method that will enable us to reveal the fine interplay of selection rules at different levels of 
word length scales and time scales not only in musical texts but also in the broader area of 
symbolic dynamics in complex system studies.

\section*{Acknowledgments}
The authors acknowledge the state-targeted program ``Center of Excellence for Fundamental 
and Applied Physics'' (BR05236454) by the Ministry of Education and Science of the Republic of 
Kazakhstan and. T.O. acknowledges the ORAU grant entitled ``Casimir light as a probe of vacuum 
fluctuation simplification'' with PN 17098. 


\addtocontents{toc}{\protect\vspace*{\baselineskip}}


\newpage

\begin{thebibliography}{1}

\bibitem{Piggott} Piggott, F. T., [1893]. \emph{The Music and musical instruments of Japan}, 
(B. T. Batsford Publ.).


\bibitem{Nicolis and Nicolis 2007} Nicolis, G. \& Nicolis, C., \emph{Foundations of Complex 
	Systems: Nonlinear Dynamics, Statistical Physics, Information and Prediction}, (World Scientific, 
2017).


\bibitem{Hao} Hao, B.L. (editor), [1990]. \emph{Chaos II: a reprint selection}, (World 
Scientific).


\bibitem{Kitchens} Kitchens, B.P., [1998]. \emph{Symbolic Dynamics: Onesided, Twosided and Countable State Markov Shifts}, Springer Verlag, Berlin.


\bibitem{Nicolis1991}  Nicolis, J.S. \emph{Chaos and Information Processing: A Heuristic Outline}, 
World Scientific (1991). 


\bibitem{NicBasBook} Nicolis, G. and Basios, V. (eds.) \emph{Chaos, Information Processing and 
Paradoxical Games: The Legacy of John S Nicolis}, Worldd Scientific (2014). 


\bibitem{Boon1995} Boon, J-P. and Decroly, O., [1995]. \emph{Dynamical systems theory 
for music dynamics}, CHAOS, 5, 501.


\bibitem{Kalimeri2012} Kalimeri M., Constantoudis V., Papadimitriou C., Karamanos K., Diakonos F., 
Papageorgiou H., [2012]. \emph{Entropy analysis of wordlength series of natural language texts: 
Effects of Text Language and Genre}, International Journal of Bifurcation and Chaos, 2012 22:09, 
(doi: 10.1142/S0218127412502239).

\bibitem{Martinakova2008} Martinakova Rendeková, Z., [2008]. \emph{Regularities in musical texts 
resulted from rankfrequency distribution of pitch}, Proceedings of the 8th conference on Systems 
theory and scientific computation, ISTASC'08, pg 124-129, World Sci. \& Eng. Ac. Soc. (WSEAS).


\bibitem{Berthe1994} Berthe, V. [1994].\emph{Conditional entropy of some automatic sequences}, J. 
Phys. A: Math, Gen., 27, 7993-8006.


\bibitem{Oikonomou2007} Provata, A. and Oikonomou, Th., [2007] \emph{Power Law Exponents 
Characterizing Human DNA}, Phys. Rev. E 75, 056102.




\bibitem{Beethoven} Beethoven, L. v., \emph{Piano Sonata Op.31 No.2}, [1801-1802]. Scores and .midi 
files: \url{http://kern.humdrum.org/}

\bibitem{Yatsuhashi Kengyoo} Yatsuhashi Kengyoo, \emph{Rokudan no Shirabe}, [1614-1685]. Scores and 
.midi files: \url{http://koto.sapp.org/}


\bibitem{Yoshizawa Kengyoo II} Yoshizawa Kengyoo II, \emph{Chidori no Kyoku}, [1808-1872]. Scores 
and .midi files: \url{http://koto.sapp.org/}

\bibitem{Byrd} Byrd, D., software \textit{MIDIFile2Text}, Indiana University.
\url{http://midifile2text.sourceforge.net/index.html}

\bibitem{ebenicolisa} Ebeling, W. \& Nicolis, G., [1992]. \emph{Word frequency and entropy of 
	symbolic sequences: A dynamical perspective}, Chaos Solit. Fract. 2, 635-650.

\bibitem{world_music} Fletcher, P. [2001]. \emph{Musics in Context: A Comprehensive Survey of the 
	World's Major Musical Cultures}, Oxford University Press.

\bibitem{music_interaction} Usaburo Mabuchi, [1994]. \emph{Music cultures in interaction: cases 
	between Asia and Europe}, Academia Music.
	
\bibitem{entropa} Basios, V.  \emph{ ENTROPA: A stand-alone program in C++ for Block-Entropy 
Analysis and Symbolic Dynamics}  GitHub Code Repository, {\tt https://github.com/Alcamis/ENTROPA}, 
committed (2018).


\bibitem{Basios et al 2008} Basios, V., Forti, L. G. \& Nicolis, G., [2008]. \emph{Symbolic dynamics 
	generated by a combination of graphs}, Int. J. Bifurcation and Chaos 18, 2265-2274.


\bibitem{Basios 2008b} Basios, V. \& Mac Kernan, D., [2008]. \emph{Symbolic dynamics, coarse 
graining and the 
	monitoring of complex systems}, Int. J. of Bifurcation and Chaos, 21, 3465-3475.


\bibitem{Gaspard and Wang 1987} Gaspard, P. \& Wang, X.-J., [1987]. \emph{Sporadicity: Between 
	periodic and chaotic dynamical behaviors}, in Proc. Natl. Acad. Sci. USA.


\bibitem{Nicolis 2005} Nicolis, J.S., [2005]. \emph{Superselection rules modulating complexity: an 
	overview}, Chaos, Solitons and Fractals, 24, 1159-1163.

\bibitem{Shannon} Shannon, C.E and Weaver, W., [1969] \emph{Mathematical Theory of Communication}, 
University of Illinois Press.

\bibitem{Gaspard 1998} Gaspard, P., [1998]. \emph{Chaos, Scattering and Statistical Mechanics}, 
(Cambridge University Press).

\bibitem{Applebaum 1996} Applebaum, D., [1996]. \emph{Information and Probability: an Integrated 
	Approach}, Cambridge University Press.





\bibitem{Bach-Recurrence} Moore, J.M., Corr{\^e}a, D.C. and Small, M., [2018].
\emph{Is Bach{\rq}s brain a Markov chain? Recurrence quantification to assess 
Markov order for short, symbolic, musical compositions}, Chaos: An Interdisciplinary Journal of 
Nonlinear Science", 28, 8, pg. 085715, (doi:10.1063/1.5024814).







\end{thebibliography}





\end{document}